\documentclass[aps,preprint,prc, showkeys,nofootinbib]{revtex4}
\usepackage{amsmath,epsfig,xcolor,soul,bm,amssymb,multirow,hyperref,xfrac,bbold} 
\newcommand{\sgn}{\mathop{\mathrm{sgn}}}
\newcommand{\BB}{$\bm B$ }
\newcommand{\HH}{$\bm H$ }
\newcommand{\MM}{$\bm M$ }
\newcommand{\bb}{$B$ }

\newcommand{\mm}{$M$ }
\usepackage{changes}
\definechangesauthor[name={JD}, color=red]{JD}
\begin{document}

\title{The spin-polarized ferromagnetic state of a cold Fermi gas} 

\author{J.P.W. Diener}
\email{dienerj@biust.ac.bw}
\affiliation{Botswana International University of Science and Technology (BIUST), Palapye, Botswana}
\affiliation{Institute of Theoretical Physics, University of Stellenbosch, Stellenbosch 7600, South Africa}
\author{F.G. Scholtz}%
\email{fgs@sun.ac.za}
\affiliation{National Institute for Theoretical Physics (NITheP), Stellenbosch 7600, South Africa}
\affiliation{Institute of Theoretical Physics, University of Stellenbosch, Stellenbosch 7600, South Africa}
\date{\today}

\begin{abstract}
The spin-polarized ferromagnetic state of a cold Fermi gas is investigated for interacting and non-interacting charge-neutral and $\beta$-equilibrated gases.  The standard minimal couplings between the magnetic field and the fermions' charges and magnetic dipole moments define the fermions' interaction with the magnetic field.  Assuming a variable coupling strength between the magnetic field and the fermion (baryon) dipole moments, it is shown that a ferromagnetized state can be achieved that corresponds to a lower energy spin-polarized state with a magnetic field entirely due to the gas's magnetic response.  We find that, depending on the density, a very large increase in the baryon dipole moments is needed to achieve this ferromagnetized state.  While the required increase seems unlikely, the induced magnetic field is of the order $\approx10^{17}$ gauss. Furthermore, while externally magnetized Fermi gases have an anisotropic pressure, the pressure of the ferromagnetized gas is completely isotropic and in the thermodynamically preferred magnetized state. 
\end{abstract}

\maketitle 
%
%
%
%
%
\section{Introduction}
The study of a highly magnetized Fermi gas, whether interacting or non-interacting, finds application at all length scales: from relativistic heavy ion collisions to neutron stars. Neutron stars are compact astrophysical objects consisting of matter at densities beyond that of nuclear matter.  These stars can also possess extremely strong magnetic fields and the magnetic field can play a key part in the observational behaviour through its impact on the star's equations of state (EoS).  Unfortunately the EoS is currently impossible to be determine exactly since the neutron star interior cannot yet be directly probed.  Models for the neutron star EoS are based on that of dense matter systems with input from high density nuclear experiments and neutron star observations \cite{ARLat, AROz, ARHe}.\\
\\
The magnetic field breaks the spherical symmetry of the Fermi surface and thus introduces an anisotropy in the pressure of the gas \cite{Bland1982}.  Whether and how this influences the EoS has been a cause of debate \cite{Ferrer2010,CommentFerrer2012,CommentFerrerFerrer2012}. The main point of contention seems to stem from the treatment of the different contributions to the magnetic field, i.e. from the field $H$ and the magnetization $M$ (see Section \ref{subsec:magf} for further definitions of these quantities).   However, these influences on the EoS were clarified in Ref. \cite{Chat2015}, where the full derivations of most magnetic properties and their relations to one another are given.  \\
\\
The magnetic response of the system in terms of a sustained, spin-polarized (ferromagnetic) state has been investigated at supranuclear densities since the late sixties \cite{BN,Silver,Lee}. While these initial studies did find a spin-polarized state in dense neutron matter likely, the magnetic field's contribution to the EoS was not included.  Currently there does not seem to be agreement over whether such a phase transition is possible in fermionic matter.  Most recently, it was argued that such a phase transition is excluded by the observed maximum mass of neutron stars \cite{Schenk2020}. However, by the authors own admission, the influence of the magnetic field was not included in their study, which could have an impact if there is a net magnetization of the matter \cite{Schenk2020}.\\
\\
In this work the computation of the magnetic field is treated self-consistently, i.e. as the solution of a self-consistency equation for a given value of $H$.  For $H=0$, the magnetic field is the result of a spin-polarized state of the gas only.  This is similar to the system that has been investigated in Refs. \cite{Brod00,Lee}, but in these studies the magnetic field was not self-consistently generated. We show that in the case of the self-consistent ferromagnetized state the pressure of the system is completely isotropic. \\
\\
We describe systems of both neutral and charged fermions (in particular baryons) interacting with a magnetic field under the condition of charge-neutrality as well as $\beta$-equilibrium.  These conditions are usually considered appropriate to describe dense nuclear matter and neutron star matter \cite{ARLat}. To incorporate neutral fermions in magnetized matter system Broderick {\em et al.}\ \cite{Brod00} included a coupling between the magnetic dipole moment and the magnetic field. It was done to take into account the higher-order contributions to the baryon dipole moments due to their internal structure. This coupling to the electromagnetic field tensor $F^{\mu\nu}$ is referred to as the ``anomalous magnetic moment'' or ``AMM''-coupling\footnote{As was noted in \cite{nuc}, we think that this term is somewhat misleading since in this case this coupling accounts for the substructure's contribution to the baryon magnetic dipole moment and not quantum corrections to it. However, we continue to use this term since it has been widely used in the literature.}. \\
\\
Charged fermions is coupled in the normal manner to the electromagnetic potential, $A^\mu$, as well as through the AMM-coupling.  The AMM-coupling for leptons are excluded since its impact has been found to be negligible \cite{mao,mao2} and it is not expected to vary with density since leptons are fundamental particles.\\
\\
The way we investigate the ferromagnetic (spin-polarized) state is to consider the variation of the AMM-coupling of protons and neutrons (fermions, in particular baryons). At the free values of these AMM-couplings the magnetization of the baryon gas is very small \cite{Brod00} and the influence on the EoS insignificant \cite{Ferrer2015AMMinsig}.  However, baryons are composite particles and at high densities its internal dynamics may be altered, thus changing the coupling strength of its AMM-coupling. Some indication of the medium effect on the magnetic dipole moment could be gleamed from that of copper and nickel isotopes.  Copper has one proton outside the $Z=28$ proton shell, while $^{56}$Ni is doubly magic.  Experimental investigations (summarized in \cite{Cudipole}) of these isotopes' magnetic moments  shows considerable variation.  Copper, with an even number of valence neutrons, has shown an increase of about 50$\%$ over a mass number range of 10 \cite{berryman}.  Since the Cu neutrons are paired, the change in the dipole moment can be interpreted as the medium effects on the single proton's magnetic dipole moment.  \\
\\
As we will report the proton and neutron AMM-couplings need to increase by a factor of between $30$ to $40$ in order for the system to be in a stable spin-polarized (ferromagnetic) state.  Since there is currently not any direct experimental evidence for such a state to exist we do not claim that these calculated values necessarily reflect the actual possible variation of the baryon AMM-coupling with density nor the exact mechanism or system which could achieve a stable spin-polarized state. However, it is important to understand the properties of a spin polarized state, e.g. the equation of state, if one would hope to find any experimental evidence for the existence of such a state.  The latter is our main motivation for the phenomenological exploration of this phase and we will present a phenomenological calculation of the value of the baryon AMM-coupling which would present an energetically favourable, lower energy state. 
	We do this for both interacting and non-interacting Fermi gases in order to establish general characteristics of such a system. \\
\\
This work will build on the already established body of work in terms of the influence of the magnetic field on fermion gases, see \cite{Brod00,Ferrer2010,Chat2015,Canuto3,Strick,Ferrer2019,diener2020} and references therein.  Most elements that we discuss have been derived elsewhere. Our contribution is to consider the conditions for a thermodynamically stable spin polarized ferromagnetic state in a cold (zero temperature) Fermi gas \cite{Lee,Canuto3} and explore the implication of such a phase transition in terms of the pressure of the Fermi gas. Our model could be applied to magnetized system where the magnetic field must be internal to the system.  While, at best, only applicable to a part of a macroscopic system such as a neutron star, the implication of an isotropic pressure in such a magnetized state could be significant.   In Section \ref{sec:ferro} we apply our description of the ferromagnetic state to a simple model of charge-neutral and $\beta$-equilibrated matter to investigate the possible existence of a phase transition to the spin polarized ferromagnetic state. The results and discussion are presented in sections \ref{sec:res} and \ref{sec:disc}.\\
\\
The signature of the Minkowski metric, $n^{\mu\nu}$, used is $(+,-,-,-)$.  Natural units, $\hbar=c=1$, are employed in conjunction with the Heaviside-Lorentz unit convention where the permittivity of free space is $\epsilon_0 =1$, which differs by a factor of $(4\pi)^{-1}$ from Gaussian units\footnote{In Gaussian units $\epsilon_0 =(4\pi)^{-1}$.  For more details on the Heaviside-Lorentz units, please see \cite{Strick} or \cite{dienerPhD}.}.

\section{Formalism}\label{sec:form}
A magnetized Fermi gas is described by the following Lagrangian
\begin{eqnarray}
	{\cal L}&=&
	\bar{\psi}{ (x)}
	\left[
		i\gamma^{\mu}\partial_{\mu}-q\gamma^\mu A_\mu
		-\frac{g}{2}F^{\mu\nu}\sigma_{\mu\nu}-m
	\right]\psi{ (x)}
	-\frac{1}{4}F^{\mu\nu}F_{\mu\nu}	\label{calL}
\end{eqnarray}
where 
$\psi$ is the fermion field operator, $F_{\mu\nu} = \partial_\mu A_\nu{ (x)}- \partial_\nu A_\mu{ (x)}$ is the electromagnetic field tensor, and $\sigma^{\mu\nu} = \frac{i}{2}\left[\gamma^\mu,\gamma^\nu\right]$ are the generators of the Lorentz group. In this paper we will use convention of Ref. \cite{itzykson} for the $\gamma$-matrices (for the precise expressions of quantities and their units in these conventions please see Ref. \cite{dienerPhD}). \\
\\
The thermodynamic quantities of the system described by Eq. (\ref{calL}) have already been derived and discussed in \cite{Ferrer2010,Chat2015,Strick} and are given in terms of the grand potential $\Omega$ of the system.  For temperature $T$, volume $V$, and fermion chemical potential $\mu$ 
\begin{eqnarray}
\Omega(B,\mu, T) = -\frac{\mbox{ln}\, {\cal Z}}{V \beta},
\end{eqnarray}
with $\beta =(k_B T)^{-1}$ the inverse temperature times Boltzmann's constant, and $ \cal Z$ the grand canonical partition function of the system.  The various thermodynamic relations are \cite{Bland1982,kapgale} 
\begin{eqnarray}
\frac{S}{V} = -\left(\frac{\partial\Omega}{\partial T}\right)_{\mu,B},
\ \rho = \frac{N}{V} = -\left(\frac{\partial\Omega}{\partial \mu}\right)_{B,T}.
\end{eqnarray}
\subsection{Magnetic field nomenclature}\label{subsec:magf}
Using the Heaviside-Lorentz unit convention the magnetic field will be described in terms of the relationship
\begin{eqnarray}\label{HBM}
	{\bm H} = {\bm B} - {\bm M}\mbox{ or }{\bm B} = {\bm H }+ {\bm M},
\end{eqnarray}
where \MM is the magnetization \cite{Brod00,Bland1982,Lee} or induced magnetic moment \cite{Canuto3} of the Fermi gas in response to the magnetic field. Unfortunately, there is little consensus in the literature with regards to the naming convention of \HH and $\bm B$.\\
\\
In some instances \HH is described as the ``magnetic field'' \cite{Bland1982}, the ``external applied magnetic field'' \cite{Ferrer2010}, the ``magnetic field due to the free currents'' \cite{Lee}, the ``magnetic field strength'' \cite{Brod00}, or the ``impressed magnetic field'' \cite{Canuto3} .  While \BB is given as the ``magnetic flux density'' \cite{Bland1982}, the ``magnetic induction'' \cite{Lee}, the ``resulting magnetic field'' \cite{Canuto3}, or just the ``magnetic field'' \cite{Strick}.  However, all agree on the relationship between \HH and \BB given in Eq. (\ref{HBM}).  Physically \HH is the magnetic field that is externally applied to the system, while \BB is the magnetic field experienced by the fermions.  Thus, from Eq. (\ref{HBM}), \BB also includes the collective magnetic response of the gas and thus \BB would be the magnetic field experienced by a single fermion. Hence \BB defines part of the fermion single particle energy as well as the cyclotron frequency of the charged fermions, $\omega_c=qB/2m$ \cite{Bland1982}.\\
\\
To avoid any confusion we will follow the convention of \cite{griffiths} and refer to \HH  simply as ``$\bm H$'' and \BB as the ``magnetic field''.  
\subsection{Gauge field, $A^\mu$}\label{subsec:Amu}
All thermodynamic quantities of interest are related to the energy-momentum tensor, $T^{\mu\nu}$.  Without loss of generalisation, the simplest way to derive $T^{\mu\nu}$ is in the rest frame of the gas/fluid.  This implies that, assuming a perfect conductor, the electric field in the gas vanishes, leaving only the magnetic field \cite{Chat2015}.\\
\\
Without further loss of generalisation, \BB is chosen to be in the $z$-direction, ${\bm B}=B{\hat{z}}$.  Correspondingly we will choose $A^{\mu}$ as
\begin{eqnarray}
A^\mu=(0,0,Bx,0)\label{Amu},
\end{eqnarray}
and this choice simplifies Eq. (\ref{calL}) to
\begin{subequations}\label{LLB}
\begin{eqnarray}
	{\cal L}&=\bar{\psi}{ (x)}
				\left[
					i\gamma^{\mu}\partial_{\mu}-q\gamma^\mu A_\mu
					+ g\bm\Sigma \!\cdot\!\bm B-m
				\right]\psi{ (x)}-\frac{1}{2}B^2\\
				&=\bar{\psi}{ (x)}
				\left[
				i\gamma^{\mu}\partial_{\mu}-q\gamma^\mu A_\mu
				+ g\Sigma_z B-m
				\right]\psi{ (x)}-\frac{1}{2}B^2	\label{calLB}
\end{eqnarray}
\end{subequations}
where $\bm \Sigma = \mathbb{1}_2\otimes\bm\sigma$ with $\mathbb{1}_2$ the $2\times2$ identity matrix and $\bm \sigma$ the Pauli matrices \cite{dienerPhD}. Since $ \bm\nabla\cdot\bm B=0 $ from this point onwards we will express Eq.(\ref{HBM}) as simply  $B=H+M$.\\
\\
 The spectrum of the Hamiltonian of Lagrangian (\ref{calLB}) for neutral fermions is \cite{Brod00,dienerPhD}
\begin{eqnarray}
\omega({\bm k},\lambda)
&=& \sqrt{\left(\sqrt{k_{\bot}^2+{m}^2}+\lambda g_b B\right)^2+k_{z}^2} \label{singlepatN},
\end{eqnarray}
while for fermions with charge $q$ 
\begin{eqnarray}
\omega(k_z,\lambda,n)=\sqrt{k_z^2+\left(\sqrt{{m}^2+2\,\alpha\,q B n}+\lambda  g B\right)^2}\label{singlepatP},
\end{eqnarray}
where
\begin{itemize}
	\item $\lambda = \pm 1$ distinguishes the different orientations of the fermion dipole moment, 
	\item $k_{\bot}^2 = k_{x}^2+k_{y}^2$ the momenta perpendicular to $\bm B$,
	\item $k_z$ is the momentum in the $z$-direction,
	\item $\alpha =\mbox{sgn}(qB)$, and
	\item $n$ is an integer labelling the Landau levels, where 
	\begin{eqnarray}\label{llcon}
	n=\left(n'+\frac{1}{2}-\alpha\,\frac{\lambda }{2}\right)\mbox{ with } n'=0,1,2,3...
	\end{eqnarray}
\end{itemize}
\subsection{Grand potential $\Omega$, pressure, and magnetization}\label{subsec:Omega}
For the zero temperature magnetized gas, there are two contributions to the grand potential $\Omega$: that of the matter (fermions) and that of the electromagnetic field.  To make this explicit we define
\begin{subequations}
		\begin{eqnarray}\label{Ofulla}
		\Omega(B,\mu) &=& \Omega_f(B,\mu)+\Omega_{EM}(B)\\
		&\equiv& \sum_\lambda \int\frac{d^3 k}{(2\pi)^3}
		\big\{
		-\omega+
		\left(
		\omega-\mu
		\right)
		\,\Theta\!
		\left[
		\,\omega-\mu
		\right]
		\big\}
		+\frac{B^2}{2}.\label{Ofullb}
	\end{eqnarray}
\end{subequations}
$\Omega_f(B,\mu)$ is the contribution of the fermions \cite{Ferrer2010,Strick,diener2020}, $\Omega_{EM}(B)=\frac{B^2}{2}$ the free magnetic field contribution to $\Omega$, and $\Theta$ is the Heaviside step function. The first term in the integrand is the zero point energy and refers to the contribution of the vacuum.  Its influence on magnetized neutron matter has been investigated recently \cite{diener2020}, but for the purposes of our discussion here we will ignore the zero point energy and defer its possible influence to a future investigation. \\
\\
For charged fermions the integral over $d^3k$ needs be replaced by
\begin{eqnarray}\label{d3k}
\sum_\lambda \int\frac{d^3 k}{(2\pi)^3}\rightarrow \sum_{\lambda,n}\frac{|q B|}{4\pi^2}\int
\end{eqnarray}
to account for the discrete nature of the $n$ filled Landau levels \cite{Brod00,dienerPhD}.\\
\\
The thermodynamic pressure in the system is given by $P=-\Omega$.  Since these are scalars, there is no anisotropy in the (thermodynamic) pressure \cite{Bland1982,Chat2015}. However, there have been considerable discussion regarding the influence of the magnetic field on the pressure, since the magnetic field breaks the spherical symmetry of the Fermi surface \cite{Bland1982,Ferrer2010,CommentFerrer2012,CommentFerrerFerrer2012}. According to \cite{Chat2015} the confusion arises since, while the thermodynamic pressure is isotropic, the energy-momentum tensor is anisotropic. The presence of anisotropic shear stresses in the energy momentum tensor is inherent to the configuration of the magnetic field and not due to the magnetic response of the gas.   Since the equation of state of the magnetized system is the quantity of interest in describing the magnetized gas, the effect of the magnetic field on the energy-momentum tensor cannot be ignored and the pressure is considered as anisotropic with a possible difference in the direction parallel and perpendicular to the magnetic field \cite{Chat2015}.  \\
\\
The longitudinal $P_{\parallel}$ and transverse $P_{\perp}$ pressures are given by \cite{Bland1982,Ferrer2010,Strick} 
\begin{subequations}\label{presO}
	\begin{eqnarray}
		P_{\parallel} &=& -\Omega = -\Omega_f-\frac{1}{2}{B}^2\\
		P_{\perp} &=& -\Omega_f-{B M}+\frac{1}{2}{B}^2 = P_{\parallel}+B(B-M).
	\end{eqnarray}
\end{subequations}
The difference in the sign of the free electromagnetic field's contribution to $P_{\parallel}$ and $P_{\perp}$ ($\pm\frac{B^2}{2}$) is due to free electromagnetic energy momentum tensor and independent of the source of the magnetic field  \cite{Strick}. Hence the anisotropy between $P_{\parallel}$ and $P_{\perp}$ is
\begin{eqnarray}\label{iso}
	P_{\perp}-P_{\parallel}  =  B(B-M)=BH.
\end{eqnarray}
Finally,  the magnetization of the Fermi gas $M$ is given in terms of $\Omega_f$ by \cite{Bland1982}
\begin{equation}\label{M}
	M=-\left(\frac{\partial\Omega_f}{\partial B}\right)_{\mu,T}.
\end{equation}
\subsection{The $H = 0$ field configuration }\label{subsec:H0}
It is important to realise that $H$ (\ref{HBM}) can be zero, while $M$, and thus $B$, can be non-zero. This is because $H$ is generated externally by so-called free currents.   In a para- or diamagnetic system, the system responds to this external perturbation, $H$, through a non-zero magnetization directed parallel or anti-parallel to $H$, respectively, and when $H$ vanishes the magnetization also vanishes. In a linear approximation the proportionality constant is just the para- or diamagnetic susceptibility.  In a ferromagnetic system, however, $H$ can vanish without $M$ vanishing.  In this case the magnetic field is therefore generated completely internally.  The latter is the case that interest us here.  In what follows we therefore take $H=0$ and consider if $M=B$ can be non-zero. \\
\\
From (\ref{iso}) is it clear that $H=0$ with $M=B\neq0$ defines an isotropic pressure in a potentially highly magnetized system.  This is contrary to externally magnetized systems where $B=H\neq0$, which could be highly anistropic \cite{isayev}. \\
\\
To determine $M=B$ when $H=0$ we consider the various thermodynamic quantities of such a system.  The first quantity is the grand canonical partition function, $\Omega\left(B,\mu\right)$ at zero temperature given by (\ref{Ofullb}).  As indicated, this is a function of $\mu$ and $B$.  From this we have the thermodynamic identities
\begin{equation}
\label{omthermrel}
H=\left(\frac{\partial\Omega}{\partial B}\right)_{\mu,T},\quad M=-\left(\frac{\partial\Omega_f}{\partial B}\right)_{\mu,T}, \quad \rho=-\left(\frac{\partial\Omega}{\partial\mu}\right)_{B,T}.
\end{equation}
Next we make a Legendre transformation to the energy density, which is a function of $\rho$ and $B$:
\begin{equation}
\epsilon\left(B,\rho\right)=\Omega\left(B,\mu\right)+\mu\rho.
\end{equation}
By subtracting $\frac{B^2}{2}$ on both sides, we can also write the fermion energy density
\begin{equation}
\epsilon_f\left(B,\rho\right)=\Omega_f\left(B,\mu\right)+\mu\rho.
\end{equation}
We then have the thermodynamic identities
\begin{equation}
\label{edens}
H=\left(\frac{\partial\epsilon}{\partial B}\right)_{\rho,T},\quad M=-\left(\frac{\partial\epsilon_f}{\partial B}\right)_{\rho,T}, \quad \mu=\left(\frac{\partial\epsilon}{\partial\rho}\right)_{B,T}.
\end{equation} 
Finally, we introduce the Gibbs free energy density, which is a function of $\rho$ and $P$
\begin{equation}
G\left(\rho,P\right)=\epsilon\left(B,\rho\right)+P=\mu\left(\rho,P\right)\rho,
\end{equation}
where we used $P=-\Omega$.  Note that since $\rho$ and $P$ are intensive, $\mu$ can be a function of both.  We have the thermodynamic identity
\begin{equation}
\mu=\left(\frac{\partial G}{\partial\rho}\right)_{P,T}.
\end{equation}
Although $G$ has no explicit dependence on $B$, it can have an implicit dependence via the dependence of $\rho$ and $P$ on $B$.  In the case of constant density and temperature, we then have
\begin{equation}
H=-\left(\frac{dG}{d B}\right)_{\rho,T}.
\end{equation}
At $H=0$, we have the conditions
\begin{equation}
\label{ferromcon}
\left(\frac{\partial\Omega}{\partial B}\right)_{\mu,T}=\left(\frac{\partial\epsilon}{\partial B}\right)_{\rho,T}=-\left(\frac{dG}{d B}\right)_{\rho,T}=0.
\end{equation}
This is the self-consistency equation that needs to be solved for $M=B$ and the existence of a non-trivial solution determines if a ferromagnetic phase exists or not.  In particular, note that (\ref{ferromcon}) corresponds to minimising the Gibbs free energy with respect to $M$ at fixed density and temperature, thus also the thermodynamical preferred state.\\
\\
The explicit form of this self-consistency equation for $M$ can be derived from (\ref{edens}) and (\ref{Ofullb})\cite{Brod00,dienerPhD}
\begin{eqnarray}\label{Meps}
	M=B=  -\frac{\partial}{\partial B}\sum_\lambda \int\frac{d^3 k}{(2\pi)^3}
		\omega
		\,\Theta\!
		\left[
		\,\omega-\mu
		\right]
\end{eqnarray}
\section{Ferromagnetic Fermi Gas}\label{sec:ferro}
To be able to compare results for the interacting and non-interacting Fermi gasses, we restrict our description of ferromagnetism to a gas of neutrons, protons, electrons, and muons.  At supranuclear densities the Coulomb repulsion could disrupt the system, thus the interacting gas will be investigated under the condition of charge-neutrality and $\beta$-equilibrium.  Electrons and muons are thus also included. For full details of this interacting model and the associated calculations, see Ref. \cite{dienerPhD}.\\
\\
To investigate the $H=0$ interacting Fermi gas, we consider relativistic mean-field (RMF) models for nuclear matter.  These models are based on a quantum field theory where protons and neutrons interact via the exchange of various mesons.  These models have been used to describe nuclei and nuclear matter, as well as the neutron star equation of state \cite{recentprogQHD}. We consider different parametrizations for the meson coupling constants, NL3 \cite{NL3}, FSU \cite{FSU1}, and FSU2 \cite{FSU2}, where the meson couplings have been fitted to different nuclear and/or neutron star properties. \\
\\
Under normal conditions the proton and neutron dipole moments, in units of the nuclear magneton $\mu_\mathrm{N}$, are $g_\mathrm{p}=2.793\,\mu_\mathrm{N}$ and $g_\mathrm{n}=-1.913\,\mu_\mathrm{N}$ respectively \cite{PDGmuon}. In \cite{dienerPhD} it is shown that to reproduce these values of the dipole moments, the AMM-couplings $g_\mathrm{n}$ and $g_\mathrm{p}$ must be equal to
	\begin{eqnarray}
		g_\mathrm{p}=-\frac{0.793}{2}\mu_\mathrm{N}= g_\mathrm{p}^{(0)}
		\mbox{ and }
		g_\mathrm{n}=\frac{1.913}{2}\mu_\mathrm{N}=g_\mathrm{n}^{(0)}.\label{gb0}
	\end{eqnarray}
Here it should be kept in mind that the sign is fixed by the choice of the sign of the AMM-coupling in (\ref{calL}).  Furthermore it should be noted that for the proton a contribution of $2\,\mu_\mathrm{N}$ to the magnetic dipole moment comes from the standard $ q\gamma^\mu A_\mu$ coupling in (\ref{calL}) for a charged particle, which of course is absent for neutral fermions (see \cite{dienerPhD} for the full calculation). The additional contributions to the point particle, quantum uncorrected values of $g_\mathrm{p}=2\mu_\mathrm{N}$ and $g_\mathrm{n}=0$, stems from the baryons' finite size and internal charge distributions and currents. Since the baryons' charges are fixed, any increase in its dipole moments must be caused by a change in its internal dynamics. In our formalism the strength of the baryon AMM-coupling can be adjusted by a factor of $x$ by changing $g_\mathrm{n}$ and $g_\mathrm{p}$ to \cite{dienerPhD}
\begin{eqnarray}
	g_\mathrm{n}=x\frac{1.913}{2}\mu_\mathrm{N}=x g_\mathrm{n}^{(0)}\mbox{ and }
 	g_\mathrm{p}=-\frac{2.793x-2}{2}\mu_\mathrm{N} = x g_\mathrm{p}^{(0)}.\label{gb}
\end{eqnarray}\\
In addition to its density dependence, the possible isospin dependence of the baryon magnetic dipole is also not known. Since the baryons have a similar three quarks substructure, we make the simplest assumption to investigate the phase boundary; namely that both baryon dipole moments will change by the same factor. However, we do not expect the existence of the phase transition and the qualitative features of the neutron star matter equation of state to depend sensitively on the isospin dependence of the magnetic dipole moment. We should point out that by increasing the dipole moments symmetrically using (\ref{gb}), the $2:3$ ratio of the dipole moment strengths is also preserved. \\
\\
From \cite{dienerPhD} the energy density of magnetized neutron star matter, which is dependent on $B$ and the total baryon density $\rho_\mathrm{b}=\rho_\mathrm{p}+\rho_\mathrm{n}$, is 
\begin{eqnarray}\label{epsLL2}
	\epsilon_f(B,\rho_\mathrm{b}) 
		&=&\sum_{\lambda,n}\frac{|q_\mathrm{p} B|}{4\pi^2}\int \omega_\mathrm{p}(k_\mathrm{z},\lambda,n)\Theta\big[\,\omega_\mathrm{p}-\mu_\mathrm{p}\big]d k_\mathrm{z}
		+\sum_{l,\lambda,n}\frac{|q_\mathrm{l} B|}{4\pi^2}\int \omega_\mathrm{l}(k_\mathrm{z},\lambda,n)\Theta\big[\,\omega_\mathrm{l}-\mu_\mathrm{l}\big]d k_\mathrm{z} \nonumber \\
		&&+\sum_\lambda\int\frac{d{\bm k}}{(2\pi)^3}\,\omega_\mathrm{n}({\bm k},\lambda)\,\Theta[\,\omega_\mathrm{n}-\mu_\mathrm{n}] 
		+U_\mathrm{int},
\end{eqnarray}
where $U_\mathrm{int}$ are nucleon-meson interactions of the different RMF parametrizatons. 
The densities and Fermi energies of the various particles are established by imposing the condition of charge-neutrality and $\beta$-equilibrium on the system. These are $\rho_\mathrm{p}=\rho_\mathrm{e}+\rho_\mu$ and $\mu_\mathrm{n}=\mu_\mathrm{p}+\mu_\mathrm{e}$ respectively. Muons are assumed to populate the system when the electron chemical potential is larger than the muon rest mass,  $\mu_\mathrm{e} > m_\mathrm{\mu}$, after which the condition $\mu_\mu=\mu_\mathrm{e}$ is also imposed. \\
\\
For $H=0$ the ferromagnetic \bb is given by $M=B=-\frac{d}{d B}\epsilon_f(B,\rho_\mathrm{b})$.   The phase boundary was calculated by adjusting $g_\mathrm{p}$ and $g_\mathrm{n}$ using (\ref{gb}) till $M=B\neq0$.
\section{Results}\label{sec:res}
\begin{figure}
	\centering
	\includegraphics[width=.750\textwidth]{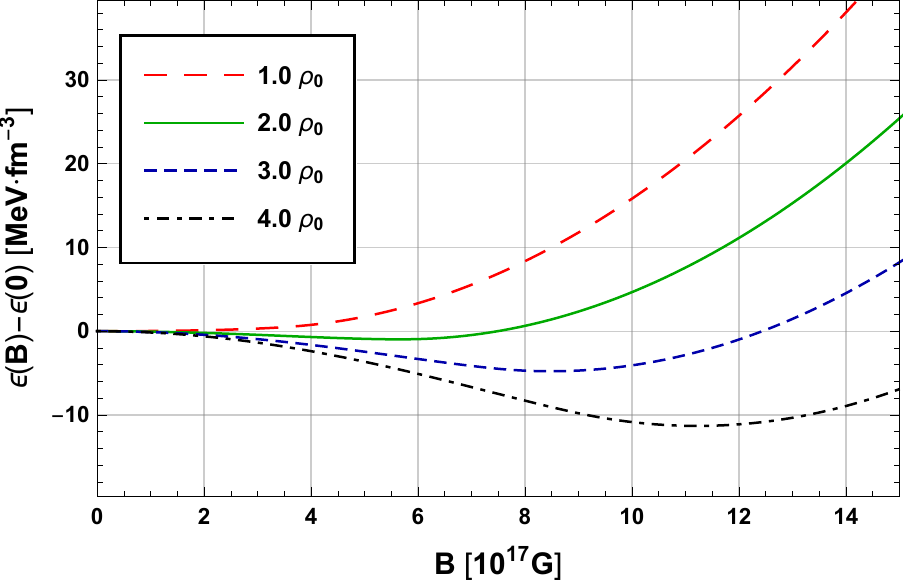}
	\caption{ The energy density relative to its value at $B=0$ i.e. $\epsilon(B,\rho_\mathrm{b})-\epsilon(0,\rho_\mathrm{b})$ of a non-interacting neutron gas as a function of $B$ for particle densities at different multiples of the nuclear saturation density, $\rho_0$.   For these plots the value of the neutron AMM-coupling is taken to be $g_\mathrm{n}=30\, g_\mathrm{n}^{(0)}$.  Since $H=0$  the magnetic field $M=B\neq0$ is given by $B$ at the energy density minimum of the various plots (\ref{ferromcon}).  }
	\label{fig:dele}
\end{figure}
\begin{figure}
	\centering
	\includegraphics[width=.750\textwidth]{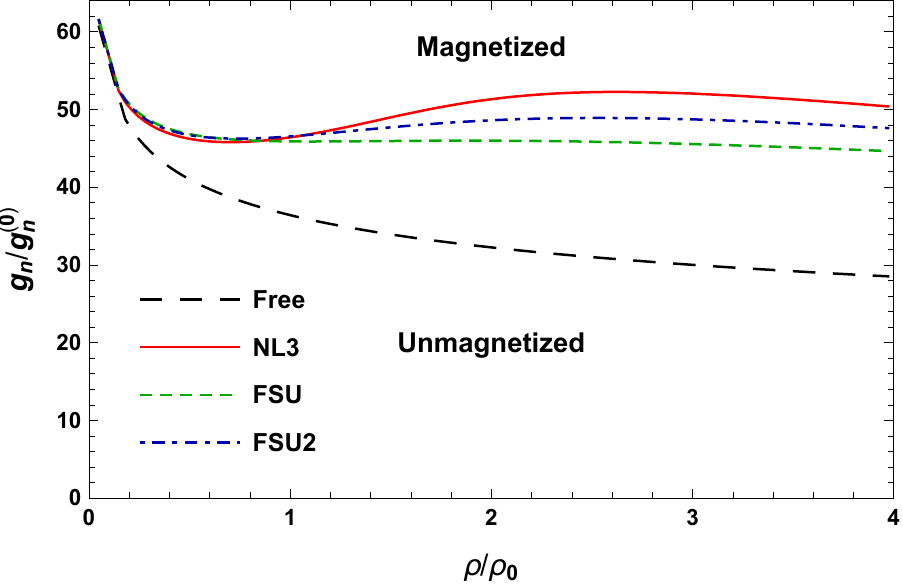}
	\caption{ Spin-polarized ferromagnetic phase boundary ($M=B\neq0$) for pure neutron matter indicated in multiples of the neutron AMM-coupling as a function of the number density (as multiples of the nuclear matter saturation density) for the indicated interacting and non-interacting systems.  }
	\label{fig:gbneu}
\end{figure}

 
Qualitatively the origin of the $M=B\neq0$ state is the magnetization induced by the asymmetric filling of the energy levels corresponding to different orientations of the fermion magnetic dipole moment.  These different orientations we denote by $\lambda=\pm 1$ in the single particle energies, Eqns (\ref{singlepatN}) and (\ref{singlepatP}). From these $m\pm g B$ are the lowest energy fermion states with an energy gap of $2\left|g B\right|$  between them. For one choice of $\lambda$  (depending on $\sgn (gB)$) lower energy states can be populated for increasing $B$, thus lowering the $\epsilon_f(B,\rho)$ while creating an asymmetric filling of different $\lambda$ states.\\
\\
Since $\mu$  and the particle's Fermi energies are independent of $\lambda$, it is the low energy assymetric filling that causes the magnetization. However, the energy gap of $2\left|g B\right|$ needs to be large enough so that the induced magnetization can sustain the ferromagnetic field, i.e. $M=B\neq0$. Also, since $\epsilon(B,\rho)\leq\epsilon(0,\rho)$, the reduction in $\epsilon_f(B,\rho)$ has to be greater than that what is gained by $\epsilon=\epsilon_f(B,\rho)+\sfrac{1}{2}B^2$ due to the contribution by $\Omega_{EM}$ for the ferromagnetic state to be stable. Thus we increased $g$ at multiples of its value at normal densities using (\ref{gb}) until the condition of $\epsilon(B,\rho)\leq\epsilon(0,\rho)$ is satisfied. Fig. \ref{fig:dele} shows the behaviour of $\epsilon(B,\rho)$ for a non-interacting neutron gas with $g_n=30\, g_\mathrm{n}^{(0)}$.
\begin{figure}
	\centering
	\includegraphics[width=.750\textwidth]{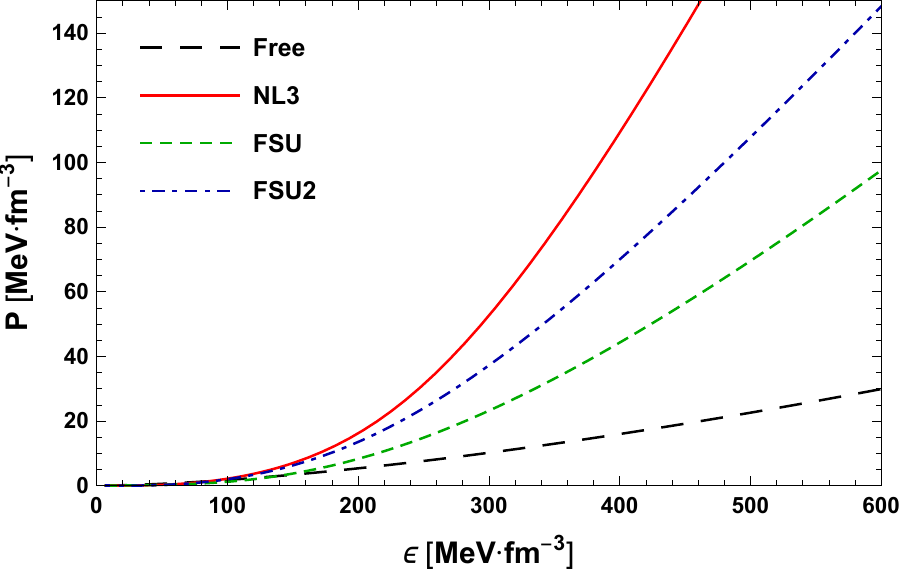}
	\caption{ Equation of state for non-interacting (``Free'') neutron gas and neutron gases interacting with the coupling strengths of the RMF parameter sets. Depending on the interactions, the corresponding (particle) density range span by $\epsilon$ as shown is from $0$ to about four times the nuclear saturation density. } 
	\label{fig:eos}
\end{figure}
\newline\newline
The EoS of the gas is defined by the relationship between $P$ and $\epsilon$. The rate at which the pressure increases is an indication of how compressible the gas is: for a ``stiff'' EoS  $P$ increases more rapidly with $\epsilon$ than for a ``soft'' EoS.  The EoS for the interacting and non-interacting gasses at the phase boundary are plotted in Fig. \ref{fig:eos}.  In a non-interacting gas of neutrons and/or protons these finite-size nucleons are treated as point particles which are very compressible.  Hence, more fermions can be contained in a volume element, thus the effect of the asymmetric filling of low energy states (magnetization) would be most pronounced for a non-interacting gas. Therefore, the non-interacting gas would  transition to the ferromagnetized state for the smallest multiple of $g$.  The phase boundaries for a gas of neutrons are shown in Fig. \ref{fig:gbneu}.
\begin{figure}
	\centering
	\includegraphics[width=.750\textwidth]{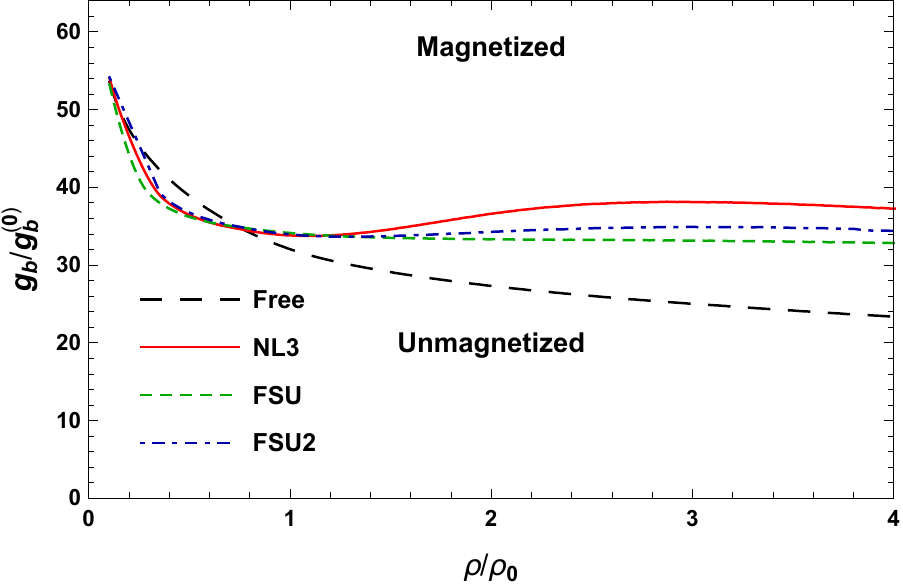}
	\caption{ Spin-polarized ferromagnetic phase boundary ($B=M\neq0$) for charge-neutral,  $\beta$-equilibrated matter in multiples of the nucleon (baryon) AMM-coupling as a function of the number density (as multiples of the nuclear matter saturation density) for the indicated interacting and non-interacting systems.  }
	\label{fig:gbsym}
\end{figure}
\newline\newline
The EoS is also softened by the inclusion of more particles.  Thus it is expected that for a charge-neutral $\beta$-equilibrated gas (with protons, neutrons and leptons) the multiples of $g$ that define the phase boundary will decrease compared to the neutron gas. This is indeed the case as is shown in Fig. \ref{fig:gbsym}. 
\begin{figure}
	\centering
	\includegraphics[width=.750\textwidth]{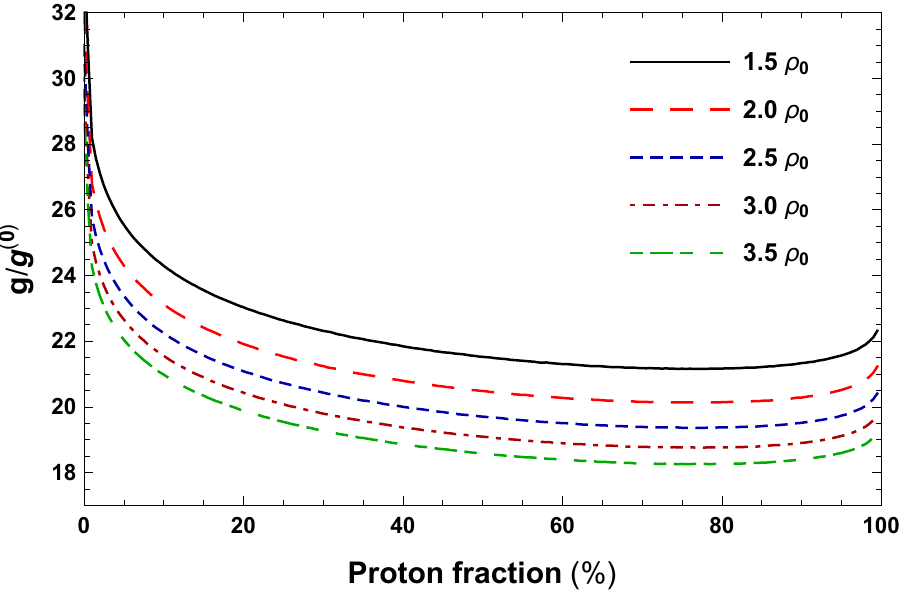}
	\caption{ Ferromagnetic phase boundary for a non-interacting charge-neutral gas for different multiples of the nuclear saturation density $\rho_0$. The ``proton fraction'' is the fraction of the proton density as a percentage of the total baryon density. }
	\label{fig:delta}
\end{figure}
\newline\newline
The isospin dependence of the baryon dipole moments is not known.  However, the phase transition occurs independently of isospin dependence.  In Fig. \ref{fig:delta} the phase boundary is shown for a charge-neutral, non-interacting gas of protons, neutrons, and leptons as a function of the proton fraction of the total baryon density.  
\begin{figure}
	\centering
	\includegraphics[width=.750\textwidth]{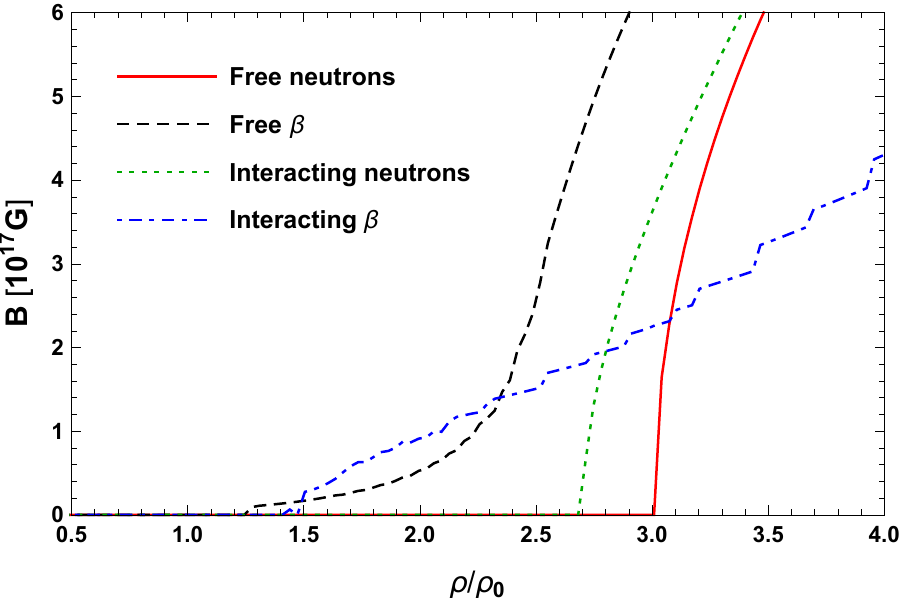}
	\caption{ The ferromagnetic fields for 
		non-interacting (free) neutron gas (``Free neutrons''), non-interacting  charge-neutral $\beta$-equilibrium (``Free $\beta$'') gas both for which $g=30 g^{(0)}$, 
		interacting (FSU parameter set) neutron gas (``Interacting neutrons'') with $g=45.75 g^{(0)}$,
		and interacting (FSU parameter set) charge-neutral $\beta$-equilibrium  gas (``Interacting  $\beta$'') with $g=33.5 g^{(0)}$. \\
		The values of $g$ were chosen to best fit the plot. The stepwise nature of the multi-fermion gases are due to the depopulation of Landau levels as $B$ increases.
	 }
	\label{fig:BB}
\end{figure}
\newline
\newline
Fig. \ref{fig:dele} indicates that, at a fixed value of the AMM-coupling, the ferromagnetic field $M=B$ increases with density.  In Fig. \ref{fig:BB} the variation of $B$ is shown as a function of density for free and interacting neutron gases, as well as $\beta$-equilibrated interacting matter for fixed values of $g$. In each instance the ferromagnetic field is of the order of $10^{17}$ gauss. \\
\\
The presence of the magnetic field also influences the EoS.  Fig. \ref{fig:ppb} shows the EoS of the same systems as in Fig. \ref{fig:BB} and compare their respective EoS with their $B=0$ counterparts. For the non-interacting gases the softening of the EoS shown is quite dramatic in the ferromagnetic state. However, it should be noted that for non-interacting gases the fermions are treated as idealised point particles which are very compressible.  Hence the softening is much more pronounced than for interacting gases where the short ranged repulsive interaction prevents extreme compression and thus the severe softening of the EoS. For the interacting gases shown in Fig \ref{fig:ppb} the softening occurs, but is almost indistinguishable from the $B=0$ case when both are plotted. 
\begin{figure}
	\centering
	\includegraphics[width=.750\textwidth]{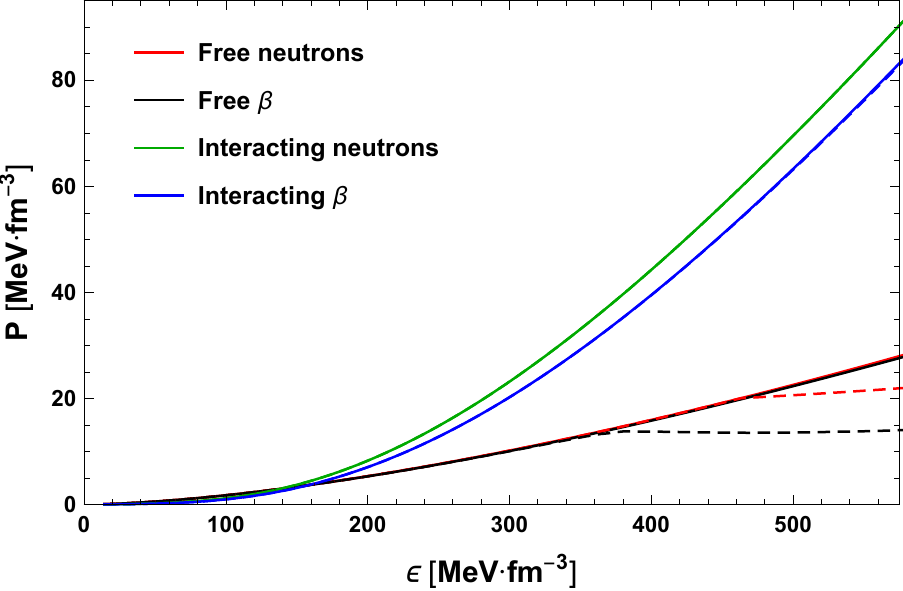}
	\caption{ The EoS for 
	non-interacting (free) neutron gas (``Free neutrons''), non-interacting  charge-neutral $\beta$-equilibrium (``Free $\beta$'') gas both for which $g=30 g^{(0)}$, 
	interacting (FSU parameter set) neutron gas (``Interacting neutrons'') with $g=45.75 g^{(0)}$,
	and interacting (FSU parameter set) charge-neutral $\beta$-equilibrium  gas (``Interacting  $\beta$'') with $g=33.5 g^{(0)}$.  
		The solid lines represent the $B=0$ values, while dashed lines give the respective $B=M\neq 0$ behaviours.  The energy densities at which the phase transitions occur are pointed out by the vertical arrows. The corresponding particle density range is the same as in Fig.\ref{fig:eos}.  }
	\label{fig:ppb}
\end{figure}
\section{Discussion}\label{sec:disc}
The conditions under which $H=0$ were investigated for charge-neutral interacting and non-interacting Fermi gases, in particular gases consisting of nucleons (protons and neutrons) and leptons (electrons and muons).  It was established that the magnetization \mm of the gas could be such that $M=B\neq0$.  However, this requires a significant increase in the strength of the nucleon AMM-coupling to the magnetic field. The strong values of the AMM-coupling needed makes it clear that the ferromagnetic phase transition would be a high energy/density effect. Its occurrence is dependent on whether this is the most effective mechanism for such a system to lower its energy.  \\
\\
The mean multiple of $g^{(0)}$ required for the phase transition is around $40$.  This represents an significant increase in the magnetic field's contribution to the fermions' energies.  As a fraction of the nucleon rest mass, $g^{(0)}$ is about $1\%$. For  $g=40g^{(0)}$, this fraction goes up to $25\%$ of the nucleon rest mass.  Thus the energy contribution to nucleon single particle energies $\omega$ (Eqns \ref{singlepatN} and \ref{singlepatP}) is of the order of the rest mass. Given the mechanism  explained in Sec. \ref{sec:res}, this increase should not be unexpected, however, physically it seems unrealistic that this would occur.  \\
\\
While still significantly large, it was demonstrated that for a softer EoS the phase transition occurs at lower multiples of $g^{(0)}$. The introduction of more fermions to the system softens the EoS.  While only nucleons were considered here, it is accepted that at higher densities various exotic particles like hyperons or even quark matter states can be populated \cite{ARLat}. \\
\\
Additionally, it was shown that the isospin dependence of the ferromagnetic phase transition favours larger proton fractions, which can decrease the AMM-coupling by more than $20\%$.  While this type of matter is not in $\beta$-equilibrium, it is charge-neutral.  While the Coulomb repulsion will destabilise any dense nuclear matter system, $\beta$-equilibrium is imposed since free neutrons have a limited lifetime \cite{ARLat}.  Hence, due to frozen in fractions, the gas might not be in $\beta$-equilibrium, which would favour the phase transition. \\ 
\\
Since our calculation is not concerned with the density dependence of the baryon AMM-coupling it cannot be determined whether a ferromagnetic phase transition will take place within a specific density range, but can only explore the characteristic of such a system. One way to better estimate how the baryon's properties might change with density could be to calculate it using chiral soliton models or the MIT-bag models. Such a calculation is one of our future aims.\\ 
\\
If the ferromagnetic phase boundary is crossed the resulting magnetic field is of the order of $10^{17}$ gauss. These field strengths are comparable to those inferred to be present in the interior of highly magnetized neutron stars known as ``magnetars'' \cite{KandK, FandR}. Based on our results the presence of a ferromagnetic phase will certainly indicate the preference for a softer EoS or multiple fermion EoS. The recent discovery of a neutron star whose mass is more than double that of our sun would appear to rule out soft equations of state. However, this star is not classified as a magnetar \cite{2msol}. Magnetars are characterized by their $X$- and/or $\gamma$-ray emissions which are indicative of their strong magnetic fields, but very little is observationally known about their masses and EoS \cite{McGill}. \\
\\
We showed that in the case of a spin-polarized ferromagnetic state with no external magnetic field ($H=0$), there is no anisotropy in the pressure. This is in contrast to magnetized Fermi gases where $H\neq0$ and large possible differences in the pressure are expected.  Unfortunately our calculation is not sophisticated enough to be applied directly to the magnetar interior. In order to do that, amongst other, the boundary conditions of the electromagnetic field and the density-dependence of the AMM-coupling needs to be incorporated. Furthermore, mechanisms to generate the long-range correlation between dipole moments necessary for a global ferromagnetic phase also have to be included.  However, based on the results in Fig. \ref{fig:ppb} for interacting gases, we do not expect a major deviation in the behaviour or the EoS of the spin-polarized ferromagnetic state compared to the unmagnetized state.  This is the focus of our future research. 
\section{Conclusion}
We investigated magnetized Fermi gases where the magnetic field is not externally applied to the gas, but internally generated in a self-consistent manner that minimizes the Gibbs free energy of the system.  We computed the ferromagnetic phase diagram for charge-neutral, $\beta$-equilibrated interacting and non-interacting Fermi gases as a function of the strength of the baryon AMM-coupling and the total baryon density. We correlated the behaviour of the phase boundaries to that of the equation of state and showed that for gases where $H=0$ no anisotropy is expected in the pressure. While our results suggests that the ferromagnetic phase could be ruled out for nucleonic matter, there are indications that the phase could be more plausible in more exotic Fermi gases/dense matter systems.  
\section{Acknowledgements}
This research is supported by the BIUST Initiation Grant No. R00047, the South African SKA project as well as the National Research Foundation of South Africa.



\begin{thebibliography}{99}
	
	\bibitem{ARLat} J.M. Lattimer,  Annual Review of Nuclear and Particle Science \textbf{62}, 485 (2012).
	\bibitem{AROz}	F.{\"{O}}zel and P.Freire, Annual Review of Astronomy and Astrophysics \textbf{54}, 401	(2016).
	\bibitem{ARHe}	K.Hebeler, J.D. Holt, J.Men{\'{e}}ndez, and A.Schwenk, Annual Review of Nuclear and Particle Science \textbf{65}, 457 (2015).
	\bibitem{Bland1982} R.D. Blandford and L. Hernquist, J. Phys. C \textbf{15}, 6233  (1982).
	\bibitem{Ferrer2010} E.J. Ferrer, V. de la Incera, J.P. Keith, I. Portillo, and P.L. Springsteen,  Phys. Rev. C \textbf{82}, 65802 (2010).
	\bibitem{CommentFerrer2012}	A.Y. Potekhin and D.G. Yakovlev. Phys. Rev. C \textbf{85}, 39801 (2012).
	\bibitem{CommentFerrerFerrer2012} E.J. Ferrer, V. de la Incera, J.P. Keith, I. Portillo, and P.L. Springsteen, Phys. Rev. C, \textbf{85}, 39802 (2012).
	\bibitem{Chat2015}	D. Chatterjee, T. Elghozi, J. Novak, and M. Oertel, Mon. Not. R. Astron. Soc. \textbf{447}, 3785 (2015).
	\bibitem{BN} D.H. Brownell and J. Callaway, Nuovo Cimento \textbf{60B}, 169 (1969).
	\bibitem{Silver} S.D. Silverstein, Phys. Rev. Lett. \textbf{23}, 139 (1969).
	\bibitem{Lee} 	H.J. Lee, V. Canuto, H-Y Chiu, and C. Chiuderi,	Phys. Rev. Lett. \textbf{23}, 390 (1969).
	\bibitem{Schenk2020} I. Tews and A. Schwenk, Astrophys. J. \textbf{892}, 14 (2020).
	\bibitem{Brod00} A. Broderick, M. Prakash, and J.M. Lattimer, Astrophys. J. \textbf{537}, 351 (2000).
	\bibitem{nuc} 	J.P.W. Diener and F.G. Scholtz,	Phys. Rev. C \textbf{87}, 65805 (2013).
	\bibitem{mao}	G.J. Mao, A. Iwamoto, and Z-X Li, Chin. J. Astron. Astrophys. \textbf{3}, 359 (2003).
	\bibitem{mao2} G.J. Mao, V.N. Kondratyev, A. Iwamoto, Z.-X. Li, X.-Z. Wu, W. Greiner, and I.N.	Mikhailov,  Chin. Phys. Lett. \textbf{20}, 1238 (2003)
	\bibitem{Ferrer2015AMMinsig} 	E.J. Ferrer, V. de la Incera, D. Manreza Paret, A. P{\'{e}}rez Mart{\'{i}}nez,	and A. Sanchez, Phys. Rev. D \textbf{91}, 85041  (2015).
	\bibitem{Cudipole} 	P. Vingerhoets, K.T. Flanagan, J. Billowes, M.L. Bissell, K. Blaum, B. Cheal, M. De	Rydt, D.H. Forest, C. Geppert, M. Honma, M. Kowalska, J. Kr{\"{a}}mer, K. Kreim, A. Krieger, R.Neugart, G.Neyens, W. N{\"{o}}rtersh{\"{a}}user, J. Papuga, T.J. 	Procter, M.M. Rajabali, R. S{\'{a}}nchez, H.H. Stroke, and D.T. Yordanov, Phys. Lett. \textbf{703}, 34  (2011).
	\bibitem{berryman} 	J.S. Berryman, Ph.D. thesis, Michigan State University (2009).
	\bibitem{Canuto3} V. Canuto and H.-Y. Chiu, Phys. Rev. \textbf{173}, 1229 (1968).
	\bibitem{Strick}	M. Strickland, V. Dexheimer, and D.P. Menezes,  Phys. Rev. D. \textbf{86}, 125032 (2012).
	\bibitem{Ferrer2019} 	E.J. Ferrer and A. Hackebill, Phys. Rev. C \textbf{99}, 65803 (2019).
	\bibitem{diener2020}  J.P.W. Diener and F.G. Scholtz,  Phys. Rev. C \textbf{101}, 035808 (2020).
	\bibitem{dienerPhD} J.P.W. Diener,  Ph.D. thesis, Stellenbosch University (2012), arXiv:1305.7346.
	\bibitem{itzykson} 	C. Itzykson and J.B. Zuber, {\em Quantum Field Theory, Dover Books on Physics}, (Dover Publications, New York, 2012).
	\bibitem{kapgale}  J.I. Kapusta and C. Gale, {\em Finite-Temperature Field Theory: Principles and Applications}, (Cambridge University Press, Cambridge 2006).
	\bibitem{griffiths} D.J. Griffiths, {\em Introduction to Electrodynamics}, 4th ed. (Cambridge University Press, Cambridge, 2017).
	\bibitem{isayev}  A.A. Isayev and J. Yang, Phys. Rev. C \textbf{84}, 65802  (2011).
	\bibitem{recentprogQHD}  B. D. Serot and J. D. Walecka, Int. J. Mod. Phys. E \textbf{6}, 515 (1997).
	\bibitem{NL3} 	G.A. Lalazissis, J. K{\"{o}}nig, and P. Ring, Phys. Rev. C \textbf{55}, 540 (1997).
	\bibitem{FSU1} 	B.G. Todd-Rutel and J. Piekarewicz, Phys. Rev. Lett. \textbf{95}, 122501 (2005).
	\bibitem{FSU2} 	W.-C. Chen and J. Piekarewicz, Phys. Rev. C \textbf{90}, 44305 (2014).
	\bibitem{PDGmuon} 	K. Nakamura and Particle Data Group, J. Phys. G, 37(7A):75021 (2010).
	\bibitem{KandK} K. Kiuchi and K. Kotake, Mon. Not. R. Astron. Soc. \textbf{385}, 1327 (2008).
	\bibitem{FandR} 	J. Frieben and L. Rezzolla, Mon. Not. R. Astron. Soc.	\textbf{427}, 3406 (2012).
	\bibitem{2msol} 	H.T. Cromartie, E. Fonseca, S.M. Ransom, P.B.  Demorest, Z. Arzoumanian, H. Blumer,	P.R. Brook, M.E. DeCesar, T. Dolch, J.A. Ellis, R.D. Ferdman, E.C. Ferrara, 	N. Garver-Daniels, P.A.  Gentile, M.L. Jones, M.T. Lam, D.R. Lorimer, R.S. Lynch, 	M.A. McLaughlin, C. Ng, D.J. Nice, T.T. Pennucci, R. Spiewak, I.H. Stairs, 	K. Stovall, J.K. Swiggum, and W.W. Zhu, Nature Astronomy, \textbf{4}, 72 (2020).
	\bibitem{McGill} S. A. Olausen and V. M. Kaspi, Astrophys. J., 	Suppl. Ser. \textbf{212}, 6 (2014).
	%
%
\end{thebibliography}
\end{document}